\documentclass[3p,fleqn,twocolumn,elsarticle-num]{elsarticle}

\usepackage{graphicx,amsmath,amssymb,epsf}

\newcommand{\vmu}{\mbox{\boldmath $\mu$}}

\newcommand{\be}{\begin{equation}}
\newcommand{\ee}{\end{equation}}

\begin{document}

\begin{frontmatter}

\title{Semiclassical theory of persistent current fluctuations in ballistic chaotic rings}

\author{Piet W. Brouwer}
\ead{brouwer@zedat.fu-berlin.de}
\address{Dahlem Center for Complex Quantum Systems and Physics Department, Freie Universit\"at Berlin, Arnimallee 14, 14195 Berlin, Germany}

\author{Jeroen Danon}
\address{Niels Bohr International Academy, and the Center for Quantum Devices,
Niels Bohr Institute, University of Copenhagen, 2100 Copenhagen, Denmark}

\date{\today}

\begin{abstract}
The persistent current in a mesoscopic ring has a Gaussian distribution with small non-Gaussian corrections. Here we report a semiclassical calculation of the leading non-Gaussian correction, which is described by the three-point correlation function. The semiclassical approach is applicable to systems in which the electron dynamics is ballistic and chaotic, and includes the dependence on the Ehrenfest time. At small but finite Ehrenfest times, the non-Gaussian fluctuations are enhanced with respect to the limit of zero Ehrenfest time. 
\end{abstract}
\end{frontmatter}

%\maketitle

\section{Introduction}
\label{sec:1}

The fact that application of a magnetic field induces an equilibrium charge current is at the basis of the Landau diamagnetic magnetic response of metals \cite{landau1930}. For conducting rings threaded by a magnetic flux, this orbital magnetic response takes the form of a current around the ring, whereas the sign of the response may be diamagnetic as well as paramagnetic \cite{hund1938}.
%The idea that a magnetic flux threading a mesoscopic normal-metal ring can cause the flow of a equilibrium current goes back to the 1930s.\cite{hund1938} 
The recognition by B\"uttiker, Imry, and Landauer that this so-called ``persistent current'' continues to exist in the presence of elastic impurity scattering \cite{buettiker1983} and, hence, should be observable in realistic metal samples, initiated a surge in theoretical and experimental work on this paradigmatic mesoscopic phenomenon in the mid 1980s and 1990s \cite{imry2002}. Two recent experiments have revived the interest in persistent currents \cite{bleszynski2009,bluhm2009,castellanos-beltran2013}. The magnitude of the measured mean square current is in excellent agreement with the original theoretical predictions for disordered metal rings \cite{cheung1989,riedel1993}. Earlier experiments had confirmed the existence of the persistent currents \cite{levy1990,chandrasekhar1991}, but a quantitative verification of the theoretical estimates was not possible.

Whereas disorder is unavoidable in metal rings, persistent currents were also investigated in semiconductor heterostructures, for which the electron motion is ballistic \cite{mailly1993}. The most pronounced difference between ballistic and disordered-diffusive rings is the possible existence of short periodic electron trajectories in the former, for which the persistent current essentially follows the behavior of ideal one-dimensional rings without potential scattering \cite{dingle1952}. Such short trajectories may dominate the magnetic response, even if the classical dynamics in the ballistic conductor is chaotic \cite{serota1992,vonoppen1993,agam1994,jalabert1996}.

An interesting case arises if the ballistic conductor has a chaotic classical dynamics, but without short periodic trajectories encircling the magnetic flux \cite{richter1998}. Examples of such a situation are, {\em e.g.}, a ballistic ring with disc-like scatterers, referred to as a ``Lorentz gas'', or a collection of chaotic cavities arranged in a ring. Without short periodic trajectories, differences between the ballistic chaotic conductor and its disordered counterpart are much more subtle, related to the ``Ehrenfest time'' $\tau_{\rm E}$ \cite{aleiner1996},
\begin{equation}
  \tau_{\rm E} = \frac{1}{\lambda} \ln k L,
  \label{eq:tauE}
\end{equation}
where $\lambda$ is the Lyapunov exponent of the classical dynamics, $k$ is the wavenumber, and $L$ a characteristic classical length scale. Being the time required for two classical trajectories a quantum separation $1/k$ apart to acquire a classical separation $L$ under the influence of the chaotic classical dynamics, $\tau_{\rm E}$ characterizes the threshold between classical-deterministic and quantum-stochastic dynamics in ballistic structures. 
Ehrenfest-time-related effects have been considered for equilibrium properties of chaotic quantum dots \cite{aleiner1997,tian2004b,brouwer2006b,waltner2010}, and for quantum transport in open systems \cite{aleiner1996,agam2000,adagideli2003,jacquod2004,tworzydlo2004,whitney2006,whitney2007,brouwer2006,brouwer2007,brouwer2007b,petitjean2009,schneider2013}, but not for persistent currents in a ring geometry. 

In the present article we report a study of the Ehrenfest-time dependence of the mesoscopic fluctuations of the persistent current in ballistic rings in which the classical electron motion is chaotic and, after appropriate coarse graining, diffusive. We consider a grand canonical ensemble, and assume that time-reversal symmetry in the ring is broken by an applied magnetic field. In a ballistic ring, mesoscopic fluctuations of the persistent current are induced by variations of the chemical potential $\mu$; no disorder average is taken. Differences between ballistic-chaotic conductors and their disordered counterparts appear through a dependence on the Ehrenfest time $\tau_{\rm E}$ for the ballistic-chaotic case, whereas $\tau_{\rm E}$ plays no role in the case of a disordered conductor. As we show below, no $\tau_{\rm E}$-dependence is found on the level of the two-point correlation function $\langle I(\phi_1) I(\phi_2) \rangle$ of the current distribution; Only the connected three-point correlation function $K(\phi_1,\phi_2,\phi_3) = \langle I(\phi_1) I(\phi_2) I(\phi_3) \rangle_{\rm c}$, which describes deviations from the Gaussian distribution, shows a dependence on the Ehrenfest time in the case of a ballistic conductor. (Here $\phi$ is the flux threading the ring, in units of the flux quantum $hc/e$; The subscript `c' refers to the `connected average', $\langle abc \rangle_{\rm c} = \langle abc \rangle - \langle ab \rangle \langle c \rangle - \langle bc \rangle \langle a \rangle - \langle ca \rangle \langle b \rangle + 2 \langle a \rangle \langle b \rangle \langle c \rangle$.) 

Below, in Sec.\ \ref{sec:2} we describe the starting point of our theoretical approach, Gutzwiller's trace formula, and the semiclassical approximation. A calculation of the two-point correlation function is presented in Sec.\ \ref{sec:3}, and the three-point correlator is discussed in Secs.\ \ref{sec:4} and \ref{sec:4a}. We conclude in Sec.\ \ref{sec:5}.

%An interesting class of ballistic conductors consists of systems with a chaotic classical dynamics. Such ballistic systems exhibit an effectively stochastic classical dynamics. In certain cases, the stochastic classical dynamics is identical to the stochastic quantum dynamics characteristic of a disordered conductor. One example of such a system is a ballistic quantum dot, which has an ergodic classical dynamics, the same as the quantum dynamics in a disordered quantum dot. Another example is the Lorentz gas, a collection of disc-like scatterers, for which the classical dynamics is diffusive, the same as the dynamics in a disordered metal. Persistent currents in classically chaotic systems was addressed theoretically,\cite{richter1998} but we are not aware of any existing experimental investigation.

%Persistent currents. Today's experiments. Experiments on diffusive metal rings and on ballistic rings. This article: intermediate case, consisting of artificially structured ballistic ring with chaotic, diffusive classical dynamics. Interest: effect of Ehrenfest time.

%Developments in semiclassics. Semiclassical approaches to persistent currents all involve diagonal approximation. New results on off-diagonal terms need to be applied!

\section{Persistent current from Gutzwiller's trace formula}
\label{sec:2}

Starting point of our calculation of the persistent current $I$ is the thermodynamic relation
\be
  I = - \frac{e}{h} \frac{\partial \Omega}{\partial \phi},
\ee
where the thermodynamic potential at temperature $T$ and chemical potential $\mu$,
\be
  \Omega = -T \int d\varepsilon \ln(1 + e^{-(\varepsilon - \mu)/T}) \nu(\varepsilon),
\ee
is expressed as an integral of the density of states $\nu(\varepsilon)$. Following previous works on persistent currents in ballistic chaotic conductors \cite{serota1992,vonoppen1993,agam1994,jalabert1996}, we use the Gutzwiller trace formula \cite{gutzwiller1990} to express the fluctuating contribution to the density of states as a sum over periodic orbits $\alpha$ on the energy shell \cite{nakamura2004},
\be
  \nu(\varepsilon) = \frac{1}{\pi \hbar} \mbox{Re}\, \sum_{\alpha} A_{\alpha} t_{\alpha}^0 e^{i {\cal S}_{\alpha}(\varepsilon)/\hbar}.
  \label{eq:nugutz}
\ee
In this expression, the label $\alpha$ represents a periodic orbit with primitive period $t_{\alpha}^0$ and period $t_{\alpha} = m t_{\alpha}^{0}$, where $m$ is the repetition number. Further ${\cal S}_{\alpha}(\varepsilon)$ is the classical action of the orbit $\alpha$ and $A_{\alpha}$ the stability amplitude of the orbit,
\be
  A_{\alpha} = [\det ((M_{\alpha}^{0})^{m} - 1) ]^{-1/2}
\ee
where $M_{\alpha}^{0}$ is the stability matrix of the primitive orbit $\alpha$ \cite{nakamura2004}.
% The stability matrix is defined in terms of the derivatives of the perpendicular phase space coordinates $q_{\perp}'=q_{\perp}(t+t_{\alpha})$ and $p_{\perp}'=p_{\perp}(t+t_{\alpha})$ with respect to the perpendicular phase space coordinates $q_{\perp}=q_{\perp}(t)$ and $p_{\perp}=p_{\perp}(t)$, 
%\be
%  M_{\alpha} = \left. \left( \begin{array}{cc} \frac{\partial q_{\perp}'}{\partial q_{\perp}} & \frac{\partial p_{\perp}'}{\partial q_{\perp}} \\  \frac{\partial q_{\perp}'}{\partial p_{\perp}} & \frac{\partial p_{\perp}'}{\partial p_{\perp}} \end{array} \right) \right|_{q_{\perp} = p_{\perp} = 0}.
%\ee
%Our notation differs from that used by Haake and others, who use $\alpha$ to label a periodic orbit together with a reference phase space point on the orbit. The version (\ref{eq:nugutz}) of the Gutzwiller trace formula appears in, {\em e.g.}, the book by Nakamura and Harayama. The advantage of carrying the factor $t_{\alpha}$ explicitly in the expression for $\nu(\varepsilon)$, is that the stability amplitude $A_{\alpha}$ is dimensionless.
%The stability matrix $M_{\alpha}$ and, hence, the stability amplitude $A_{\alpha}$ can also be defined for a non-periodic trajectory. If the classical dynamics is chaotic with Lyapunov exponent $\lambda$, one has
%\be
%  A_{\alpha} = e^{\lambda t_{\alpha}/2},
%\ee
%Hence, if a trajectory $\alpha$ is divided up into two segments $\alpha'$ and $\alpha''$, $A_{\alpha} = A_{\alpha''} A_{\alpha'}$.

We now specialize to a two-dimensional system threaded by a flux $\Phi = \phi hc/e$. Considering energies $\varepsilon$ near the chemical potential $\mu$, the action ${\cal S}_{\alpha}(\varepsilon,\phi)$ can be written
\be
  {\cal S}_{\alpha}(\varepsilon,\phi) =
  {\cal S}_{\alpha}(\mu,0) + 2 \pi \phi \hbar n_{\alpha} + (\varepsilon-\mu) t_{\alpha},
\ee
where $n_{\alpha}$ is the winding number of the trajectory $\alpha$. Below we will write $S_{\alpha}$ as short-hand notation for ${\cal S}_{\alpha}(\mu,0)$. 
Substituting the Gutzwiller trace formula for the density of states $\nu$, taking the derivative to $\phi$, and performing the integration over $\varepsilon$, one finds \cite{richter1998}
\begin{eqnarray}
  I &=&
%  - \frac{ie}{2 \pi \hbar^2} \sum_{\alpha} n_{\alpha} t_{\alpha}
%  \int^{\mu} d\varepsilon (\varepsilon - \mu)
%  \left( A_{\alpha} e^{i ({\cal S}_{\alpha} + (\varepsilon - \mu) t_{\alpha})/\hbar + 2 \pi i n \phi} - A_{\alpha}^* e^{-i ({\cal S}_{\alpha} + (\varepsilon - \mu) t_{\alpha})/\hbar - 2 \pi i n \phi} \right) \nonumber \\ &=&
  - \frac{ie}{2 \pi \hbar} \sum_{\alpha} 
  \frac{n_{\alpha} \pi T t_{\alpha}^0}{t_{\alpha} \sinh(\pi t_{\alpha} T/\hbar)}
  \\ && \mbox{} \times
  \left( A_{\alpha} e^{\frac{i}{\hbar} {\cal S}_{\alpha} + 2 \pi i n_{\alpha} \phi} - A_{\alpha}^{*} e^{-\frac{i}{\hbar} {\cal S}_{\alpha} - 2 \pi i n_{\alpha} \phi} \right).\nonumber 
\end{eqnarray}
Upon separating the current into Fourier components,
\be
  I = \sum_{n} I_n e^{2 \pi i n \phi},
\ee
with $I_n = I_{-n}^*$, one then arrives at the result
\begin{eqnarray}
  \label{eq:In}
  I_n &=& - \frac{i e n}{2 \pi \hbar} \sum_{\alpha}
  \frac{\pi T t_{\alpha}^0}{t_{\alpha} \sinh(\pi t_{\alpha} T/\hbar)}
  \\ && \nonumber \mbox{} \times
  \left( A_{\alpha} e^{i {\cal S}_{\alpha}/\hbar} \delta_{n_{\alpha},n} + A_{\alpha}^* e^{-i {\cal S}_{\alpha}/\hbar} \delta_{n_{\alpha},-n} \right).
\end{eqnarray}
%where the trajectory $\bar \alpha$ is the time-reversed of $\alpha$, so that $n_{\bar \alpha} = - n_{\alpha}$.

\section{Mean square current}
\label{sec:3}

We now calculate the mean square $\langle I_n I_{-n} \rangle$ for the case that time-reversal symmetry in the ring is broken by an applied magnetic field. The leading contribution to $\langle I_n I_{-n} \rangle$ comes from diagonal contributions,
\begin{eqnarray}
  \langle I_n I_{-n} \rangle &=& 
  \frac{2 e^2 n^2}{(2 \pi \hbar)^2}
  \\ && \nonumber \mbox{} \times
  \sum_{\alpha} \frac{(\pi T)^2 (t_{\alpha}^0/t_{\alpha})^2}{\sinh^2(\pi t_{\alpha} T/\hbar)}
  |A_{\alpha}|^2 \delta_{n_{\alpha},n}.~~~
  \label{eq:Asum}
\end{eqnarray}
The factor two in the numerator comes from the two terms in Eq.\ (\ref{eq:In}), which give equal contributions to $\langle I_n I_{-n} \rangle$.

In order to perform the trajectory sum in Eq.\ (\ref{eq:Asum}),
%In zero-dimensional systems, the trajectory sum in Eq.\ (\ref{eq:Asum}) is performed using the Hannay-Ozorio de Almeida sum rule.\cite{hannay1984} In order to deal with the ring geometry, 
we use a method proposed by Argaman, Imry, and Smilansky \cite{argaman1993}. 
%Hereto, we write Eq.\ (\ref{eq:Asum}) as
%\be
%  \langle I_n I_{-n} \rangle = \frac{e^2 n^2}{2 \pi^2}
%  \int dt \frac{(\pi T)^2}{\sinh^2(\pi t T)}
%  \left[ \sum_{\alpha} 
%  \langle |A_{\alpha}|^2 \delta_{n_{\alpha},n}
%  \delta(t-t_{\alpha}) \right].
%\ee
The summation over classical trajectories is expressed as an integral over the energy shell $Q$. Introducing a phase space coordinate $\vmu$, and denoting with $\vmu(t)$ the phase space coordinate obtained by following the classical time evolution for a time $t$, starting at $\vmu$, one has
%Following Ref.\ \onlinecite{argaman1993}, one then has
\begin{eqnarray}
  \lefteqn{\sum_{\alpha} 
  t_{\alpha}^0 |A_{\alpha}|^2 \delta_{n_{\alpha},n}
  \delta(t-t_{\alpha})} \hspace{2cm} \\ \nonumber
  &=& 
  \int_{Q} d\vmu
  \delta(\vmu(t) - \vmu)
%  \nonumber \\ && \mbox{} \times
  \delta_{n(\mbox{\scriptsize \boldmath $\mu$},t),n}, 
\end{eqnarray}
where $n(\vmu,t)$ is the number of times the trajectory starting at the phase space point $\vmu$ winds around the flux in the time $t$. The factor $t_{\alpha}^0$ arises, because each trajectory is weighted by a factor $t_{\alpha}^0$ upon performing the phase space integration \cite{nakamura2004}. Upon identifying
\be
  \delta(\vmu(t) - \vmu)
  \delta_{n(\mbox{\scriptsize \boldmath $\mu$},t),n} =
  p(\vmu,\vmu,t|n),
\ee
as the classical probability density that a particle starting at phase space point $\vmu$ is found at the same phase space point at time $t$, while having passed $n$ times around the flux, we conclude that
\begin{eqnarray}
  \langle I_n I_{-n} \rangle &=&
  \frac{e^2 n^2}{2 \pi^2 \hbar^2}
  \int dt\, \frac{(\pi T)^2}{t \sinh^2(\pi t T/\hbar)}
  \nonumber \\ && \mbox{} \times  \int d\vmu 
  p(\vmu,\vmu,t|n).
\end{eqnarray}
Here we neglected the contribution from orbit repetitions, which is a standard approximation in this field, since the non-primitive orbits at a given period are exponentially outnumbered by primitive orbits with the same period.

For a two-dimensional ring of circumference $L$ with diffusive electron dynamics, one has
\be
  p(\vmu,\vmu,t|n) =
  \frac{L}{Q}
  \frac{e^{-(n L)^2/4 D t}}{\sqrt{4 \pi D t}},
\ee
where $Q = 2 \pi \hbar \tau_{\rm H}$ is the volume of the energy shell, $\tau_{\rm H}$ being the Heisenberg time, and $D$ the classical diffusion constant. One then arrives at the result
\begin{eqnarray}
  \langle I_n I_{-n} \rangle &=&
  \frac{e^2 n^2}{2 \pi^2 \hbar^2}
  \int dt \frac{(\pi T)^2}{t \sinh^2(\pi t T/\hbar)}
  \nonumber \\ && \mbox{} \times 
  \sqrt{\frac{\tau_{L}}{4 \pi t}} 
  e^{-\tau_{L} n^2/4 t},
%  \frac{e^{-\tau_{L} n^2/4 t}}{\sqrt{4 \pi t/\tau_L}},
  \label{eq:I2}
%  \nonumber \\ &=&
%  \frac{6 e^2 D^2}{\pi^2 n^3 L^4}.
\end{eqnarray}
where
\be
  \tau_L = \frac{L^2}{D}
\ee
is the time required to diffuse around the ring.
This is the same result as what one obtains for a disordered metal ring \cite{cheung1989,riedel1993}. In the limit of zero temperature, Eq.\ (\ref{eq:I2}) simplifies to
\be
  \langle I_{n} I_{-n} \rangle =
  \frac{6 e^2}{\pi^2 n^3 \tau_L^2}.
\ee
For high temperatures, $T \gg \hbar/\tau_L$, the integration can be performed using the saddle-point method and gives
\be
  \langle I_n I_{-n} \rangle =
  \frac{2 e^2}{\hbar^2}
  |n| T^2 e^{-|n| \sqrt{2 \pi T \tau_{L}/\hbar}},
\ee
up to corrections that are small in the limit $T \gg \hbar/\tau_L$.

The main result of this section is that the two-point correlation function is the same for a ballistic chaotic ring and for a disordered metal ring, provided the coarse-grained classical dynamics in the ring is diffusive. The Ehrenfest time $\tau_{\rm E}$ of Eq.\ (\ref{eq:tauE}) has not entered into our considerations.

\section{Third cumulant: Off-diagonal contribution}
\label{sec:4}

The second moment of the current distribution could be calculated by considering diagonal terms in the trajectory sum only. For the calculation of the connected expectation value $\langle I_{n} I_{m} I_{-n-m} \rangle_{\rm c}$, the Fourier transform of which gives the connected three-point function $K(\phi_1,\phi_2,\phi_3)$, one needs to go beyond the diagonal approximation. It is at this point, that the Ehrenfest time $\tau_{\rm E}$ enters into the calculation \cite{aleiner1996}.

\begin{figure}
\begin{center}
\includegraphics{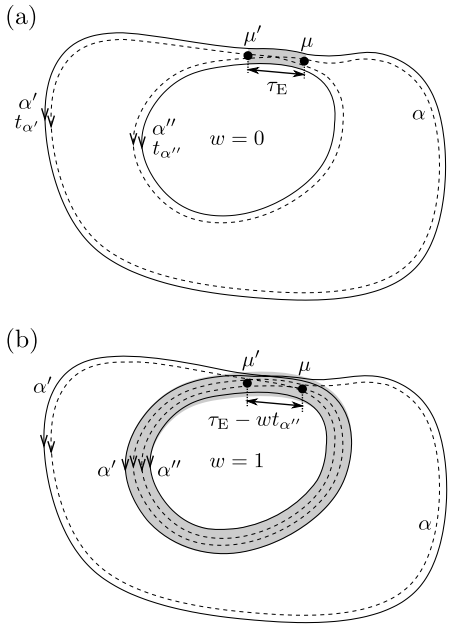}
\caption{\label{fig:eight} Trajectories $\alpha$ (dashed), $\alpha'$ (solid), and $\alpha''$ (solid) contributing to the expectation value $\langle I_{n} I_{m} I_{-n-m} \rangle$. The trajectory $\alpha$ has a ``figure eight'' structure and is piecewise equal to $\alpha'$ or $\alpha''$, up to quantum uncertainties. The differences between $\alpha$ on the one hand and $\alpha'$ and $\alpha''$ on the other hand are maximal in the $\tau_{\rm E}$-long ``encounter'' (dark shaded part) in which $\alpha$ ``switches partners''. The points $\mu$ and $\mu'$ denote the points in phase space where the self-encounter starts and ends respectively. (a) The duration of the self-encounter, $\tau_{\rm E}$, can be shorter than both $t_{\alpha'}$ and $t_{\alpha''}$. (b) The self-encounter can wrap $w$ times around the shorter of the orbits $\alpha'$ and $\alpha''$. Here $\alpha''$ is assumed to be the shorter one, and we illustrated the case of $w=1$.}
\end{center}
\end{figure}
The semiclassical calculation of a connected three-point function of the density of states (which contains all essential information for the three-point function of the persistent current) was performed by Heusler and coworkers for the case of a chaotic quantum dot \cite{heusler2007,mueller2009b}, building on previous developments of the trajectory-based semiclassical formalism by Sieber and Richter \cite{sieber2001}. Following Refs.\ \cite{heusler2007,mueller2009b}, the dominant contribution to $\langle I_{n} I_{m} I_{-n-m} \rangle_{\rm c}$ is given by a summation over ``trajectory triplets''. These trajectory triplets consist of a trajectory $\alpha$ which contains a small-angle self-encounter, so that it effectively has a ``figure-eight'' structure, see the dashed path in Fig.\ \ref{fig:eight}a. The other two trajectories $\alpha'$ and $\alpha''$ (solid paths) are different and piecewise equal to one of the loops of $\alpha$, up to a quantum uncertainty. Once the trajectory $\alpha$ is specified, the other two trajectories are uniquely determined, so that $\alpha'$ and $\alpha''$ need not be summed over separately. The periods of the three trajectories in Fig.\ \ref{fig:eight} are related as
\be
  t_{\alpha} = t_{\alpha'} + t_{\alpha''}
\ee
and for the stability amplitudes one finds
\be
  A_{\alpha} = A_{\alpha'} A_{\alpha''}.
\ee
Accounting for the various ways in which the trajectories can be combined, one obtains the expression
%\begin{widetext}
\begin{eqnarray}
  \lefteqn{ \langle I_{n} I_{m} I_{-n-m} \rangle_{\rm c}}
  \nonumber \\ &=&
  -\frac{2 i e^3 n m (n+m)}{(2 \pi)^3\hbar^3}
  \sum_{\alpha} 
  \frac{\pi T |A_{\alpha}|^2}{\sinh(\pi t_{\alpha} T/\hbar) }
  \nonumber \\ && \mbox{} \times
  \frac{(\pi T)^2\, \mbox{Re}\, e^{i ({\cal S}_{\alpha} - {\cal S}_{\alpha'} - {\cal S}_{\alpha''})/\hbar}}{\sinh(\pi t_{\alpha'} T/\hbar) \sinh(\pi t_{\alpha''} T/\hbar)}
  \nonumber \\ && \mbox{} \times
  (\delta_{n_{\alpha'},n} \delta_{n_{\alpha''},m} +
  \delta_{n_{\alpha'},n} \delta_{n_{\alpha''},-m-n} 
  \nonumber \\ && \ \ \ \mbox{} +
  \delta_{n_{\alpha'},m} \delta_{n_{\alpha''},-m-n}).
  \label{eq:Innn}
\end{eqnarray}

The summation over trajectories is now performed using the method of Refs.\ \cite{heusler2007,mueller2009b} and their extension to the systems with diffusive classical dynamics \cite{brouwer2007b}. The oscillating factor in the numerator of Eq.\ (\ref{eq:Innn}) suppresses contributions from all trajectory triplets for which the action difference $\Delta {\cal S} = {\cal S}_{\alpha} - {\cal S}_{\alpha'} - {\cal S}_{\alpha''}$ between $\alpha$ on the one hand and $\alpha'$ and $\alpha''$ on the other hand is not at most of order $\hbar$. Since the action difference is related to the duration $\tau_{\rm enc}$ of the small-angle encounter \cite{mueller2004,mueller2005}, $|\Delta {\cal S}/\hbar| \sim k L e^{-\lambda \tau_{\rm enc}}$, one finds that only trajectory triplets with $\tau_{\rm enc} = \tau_{\rm E}$ contribute to the summation. Proceeding as in Ref.\ \cite{brouwer2007b} one then finds that
%In order to ensure that the action difference between $\alpha$ on the one hand and $\alpha'$ and $\alpha''$ on the other hand is at most of order $\hbar$ --- otherwise the fast oscillations in the exponent in Eq.\ (\ref{eq:Innn}) suppress the contribution to the correlation function ---, the duration of the small-angle encounter, defined as the length of the time interval for which the phase space distance between the two nearby stretches of $\alpha$ is small enough that the classical mechanics can be linearized, must be equal to the Ehrenfest time $\tau_{\rm E}$.\cite{aleiner1996,sieber2001} The summation over trajectories, we adapt the calculation of Refs.\ \cite{heusler2007,mueller2009b} to the ring geometry. Referring to Ref.\ \cite{brouwer2007b} for details, we find that the trajectory sum is equal to
\begin{eqnarray}
  \lefteqn{\sum_{\alpha} |A_{\alpha}|^2 \cos(\Delta {\cal S}/\hbar)
  \delta_{n_{\alpha'},n} \delta_{n_{\alpha''},m}\delta(t_{\alpha'} - t_1) }
   \\ &&
  \mbox{} \times  \delta(t_{\alpha''} - t_2)
   =
  \sum_{w\geq 0}\frac{1}{\tau_{\rm H}} \frac{\partial}{\partial \tau_{\rm E}}
  F(t_1,t_2;\tau_{\rm E};w),\nonumber
\end{eqnarray}
where the function $F(t_1,t_2;\tau_{\rm E};w)$ is the properly normalized probability for the trajectory configuration to occur. It depends on the number of times $w$ the encounter winds around the shorter of the orbits $\alpha'$ and $\alpha''$,
\begin{eqnarray}
  \lefteqn{F(t_1,t_2;\tau_{\rm E};w)} \label{eq:t2_short}
  \\ &=&
  Q
  \int d\vmu d\vmu'    
  p(\vmu,\vmu',t_1-\tau_{\rm E}|n- w m) 
  \nonumber \\ && \nonumber \mbox{} \times
  p(\vmu',\vmu,\tau_{\rm E} - w t_2) 
  \nonumber \\ && \nonumber \mbox{} \times
  p(\vmu,\vmu',t_2-\tau_{\rm E} + w t_2|m)
\end{eqnarray}
if $t_1 > \tau_{\rm E}$ and $\tau_{\rm E}/(w+1) < t_2 < \tau_{\rm E}/w$, {\em i.e.}, if $t_2$ is the duration of the shorter orbit and the encounter wraps $w$ times around it,
\begin{eqnarray}
  \lefteqn{F(t_1,t_2;\tau_{\rm E};w)} \\ &=&
  Q
  \int d\vmu d\vmu'    
  p(\vmu,\vmu',t_2-\tau_{\rm E}|m-wn) \nonumber \\ && \nonumber \mbox{} \times
  p(\vmu',\vmu,\tau_{\rm E} - w t_1) \nonumber \\ && \nonumber \mbox{} \times
  p(\vmu,\vmu',t_1 -\tau_{\rm E} + w t_1|n)
\end{eqnarray}
if $t_2 > \tau_{\rm E}$ and $\tau_{\rm E}/(w+1) < t_1 < \tau_{\rm E}/w$, {\em i.e.}, if $t_1$ is the duration of the shorter orbit and the encounter wraps $w$ times around it, and
\begin{eqnarray}
  F(t_1,t_2;\tau_{\rm E};w) &=& 0
\end{eqnarray}
if both $t_1$ and $t_2$ are smaller than $\tau_{\rm E}$, in which case the figure-eight configuration of Fig.\ \ref{fig:eight} is not possible because the trajectories $\alpha'$ and $\alpha''$ are identical and $\alpha$ is a non-primitive orbit. 
Figure \ref{fig:eight}a shows an example of an orbit configuration with $w=0$; An example with $w=1$ is shown 
schematically in Fig.\ \ref{fig:eight}b.
%The orbits sketched in Fig.~\ref{fig:eight}b correspond to the %case of Eq.~(\ref{eq:t2_short}), where $\alpha''$ is the 
%shortest orbit, and the encounter (shaded part) winds once
%around $\alpha''$ ($w=1$).
%The Heisenberg time $\tau_{\rm H}$ is related to the volume of the energy shell $Q$ as $Q = 2 \pi \hbar \tau_{\rm H}$.

Denoting the distance around the ring's circumference between the phase points $\vmu$ and $\vmu'$ point by $x$, we have
\begin{eqnarray}
  p(\vmu,\vmu',t|n) &=&
  \frac{L}{Q} \frac{e^{-(n L-x)^2/4 D t}}{\sqrt{4 \pi D t}}, \nonumber \\
  p(\vmu',\vmu,t) &=&
  \frac{L}{Q} \frac{e^{-x^2/4 D t}}{\sqrt{4 \pi D t}},
\end{eqnarray}
where we do not impose a bound on $x$ to account for the possibility that the encounter itself winds around the ring.
%where we excluded the possibility that the encounter itself has a nonzero winding number. 
Substituting these explicit expressions for the probability densities, one finds
\begin{eqnarray}
  \label{eq:Fresult1}
  \lefteqn{
  F(t_1,t_2;\tau_{\rm E};w) =
  \frac{\tau_{\rm L}}{4 \pi\sqrt{\sigma(t_2,t_1;\tau_{\rm E};w)}}}
  \\ && \nonumber \mbox{} \times
  e^{- \frac{\tau_L (n^2 t_2 - 2 m n \tau_{\rm E} + m^2 (t_1 - w(w+1) t_2 + 2w \tau_{\rm E}))}{4 \sigma(t_2,t_1;\tau_{\rm E};w)}}
\end{eqnarray}
if $t_1 > \tau_{\rm E}$ and $\tau_{\rm E}/(w+1) < t_2 < \tau_{\rm E}/w$,
\begin{eqnarray}
  \label{eq:Fresult2}
  \lefteqn{
  F(t_1,t_2;\tau_{\rm E};w) =
  \frac{\tau_{\rm L}}{4 \pi\sqrt{\sigma(t_1,t_2;\tau_{\rm E};w)}}}
  \\ && \nonumber \mbox{} \times
  e^{- \frac{\tau_L (m^2 t_1 - 2 m n \tau_{\rm E} + n^2 (t_2 - w(w+1) t_1 + 2w \tau_{\rm E}))}{4 \sigma(t_1,t_2;\tau_{\rm E};w)}}
\end{eqnarray}
if $t_2 > \tau_{\rm E}$ and $\tau_{\rm E}/(w+1) < t_1 < \tau_{\rm E}/w$, and
\begin{eqnarray}
%  F(t_1,t_2;\tau_{\rm E}) &=&
%  \frac{\tau_{\rm L}e^{- \tau_L (n^2 t_2 - 2 m n \tau_{\rm E} + m^2 (t_1 - w(w+1) t_2 + 2w \tau_{\rm E}))/4 \sigma(t_2,t_1;\tau_{\rm E})}}{4 \pi\sqrt{\sigma(t_2,t_1;\tau_{\rm E})}}\ \
%  \mbox{if}\
%  t_1 > \tau_{\rm E}\
%  \mbox{and}\
%  \frac{\tau_{\rm E}}{k} < t_2 < \frac{\tau_{\rm E}}{k-1}, \nonumber \\
%  F(t_1,t_2;\tau_{\rm E}) &=&
%  \frac{\tau_{\rm L}e^{- \tau_L (m^2 t_1 - 2 m n \tau_{\rm E} + n^2 (t_2 - w(w+1) t_1 + 2w \tau_{\rm E}))/4 \sigma(t_1,t_2;\tau_{\rm E})}}{4 \pi \sqrt{\sigma(t_1,t_2;\tau_{\rm E})}}
%  \ \
%  \mbox{if}\
%  t_2 > \tau_{\rm E}\
%  \mbox{and}\
%  \frac{\tau_{\rm E}}{k} < t_1 < \frac{\tau_{\rm E}}{k-1}, \nonumber \\
  F(t_1,t_2;\tau_{\rm E};w) &=& 0
%   \ \
%  \mbox{if}
%  t_1 < \tau_{\rm E}\
%  \mbox{and}\
%  t_2 < \tau_{\rm E}.
  \label{eq:Fresult3}
\end{eqnarray}
%\end{widetext}
if $t_1 < \tau_{\rm E}$ and $t_2 < \tau_{\rm E}$.
Here
$$
  \sigma(t_1,t_2;\tau_{\rm E};w) =
  t_1 t_2 - \tau_{\rm E}^2 - w(w t_1 + t_1 - 2 \tau_{\rm E}) t_1.
$$%
We note that $F$ is continuous at $t_{1,2} = \tau_{\rm E}/w$ with $w=1,2,\ldots$.

We first perform the remaining integration over $t_1$ and $t_2$ in the limit $\tau_{\rm E} \ll \tau_{L}$. In this limit, it is sufficient to consider the case $w=0$ only, and we may take the limit $\tau_{\rm E}/\tau_{L} \to 0$ after differentiation to $\tau_{\rm E}$. We find
\begin{eqnarray}
  \lefteqn{\sum_{\alpha} |A_{\alpha}|^2 \cos(\Delta {\cal S}/\hbar)
  \delta_{n_{\alpha'},n} \delta_{n_{\alpha''},m}
  \delta(t_{\alpha'} - t_1)}
  \nonumber \\ \mbox{} \lefteqn{\times \delta(t_{\alpha''} - t_2)}
  \nonumber \\
  &=& 
%%  2 \pi \hbar
%%  \frac{\partial}{\partial \tau_{\rm E}}
%%  \frac{L^2}{Q} \int_{-\infty}^{\infty} dx
%%  \frac{e^{-x^2/4 D \tau_{\rm E} - (n L-x)^2/4 D (t_1 - \tau_{\rm E}) -
%%  (m L-x)^2/4 D (t_2 - \tau_{\rm E})^2}}{\sqrt{(4 \pi D)^{3} \tau_{\rm E} (t_1-\tau_{\rm E}) (t_2 - \tau_{\rm E})}} \nonumber \\ &=&
%  \left.
%  \frac{\tau_{L}}{4 \pi \tau_{\rm H}} 
%  \frac{\partial}{\partial \tau_{\rm E}}
%  \frac{e^{-\tau_L (m^2 t_1 + n^2 t_2 - 2 m n \tau_{\rm E})/4 (t_1 t_2 - \tau_{\rm E}^2)}}{\sqrt{t_1 t_2 - \tau_{\rm E}^2}} \right|_{\tau_{\rm E} \to 0}
%  \nonumber \\ &=&
  \frac{m n \tau_L^2}{8 \pi \tau_{\rm H}(t_1 t_2)^{3/2}} e^{-m^2 \tau_L/4 t_2 - n^2 \tau_L/4 t_1}.  
  \label{eq:triplesum}
\end{eqnarray}
Performing the remaining integrations over $t_1$ and $t_2$ in the limit of zero temperature then gives
\begin{eqnarray}
  \lefteqn{
\sum_{\alpha} \frac{|A_{\alpha}|^2 \cos(\Delta {\cal S})}{t_{\alpha} t_{\alpha'} t_{\alpha''}}
  \delta_{n_{\alpha'},n} \delta_{n_{\alpha''},m}
  }~~  && \nonumber \\ 
  &=&
  \frac{12}{\tau_{\rm H} \tau_L^2} f_{n,m} ,
\end{eqnarray}
with
\begin{eqnarray*}
  f_{n,m} &=& 
%  \frac{1}{m^2 n^2}
%  \int_0^{\infty} dx_1 dx_2
%  \frac{(x_1 x_2)^{3/2} e^{-x_1-x_2}}{m^2 x_1 + n^2 x_2} \\ &=&
  {\rm sign}(mn) \frac{|m|^2 + 4 |m| |n| + |n|^2}{m^2 n^2 (|m|+|n|)^4}.
\end{eqnarray*}
Hence, in the limit $\tau_{\rm E} \ll \tau_{L}$ and at zero temperature, one finds
\begin{eqnarray}
  \label{eq:Kresult}
  \lefteqn{\langle I_n I_m I_{-n-m} \rangle} \nonumber \\ &=&
  \langle I_{-n} I_{-m} I_{n+m} \rangle^* \nonumber \\ &=& 
  \frac{3 e^3 m n (n+m)}{i \pi^3 \tau_L^2 \tau_{\rm H}}
  \\ && \mbox{} \nonumber \times
  (f_{n,m} + f_{n,-m-n} + f_{m,-m-n}).~~~
%  [f(n,m) - f(n,m+n) - f(m,m+n)],
\end{eqnarray}
%where
%\begin{eqnarray*}
%  f(n,m) &=& 
%%  \frac{1}{m^2 n^2}
%%  \int_0^{\infty} dx_1 dx_2
%%  \frac{(x_1 x_2)^{3/2} e^{-x_1-x_2}}{m^2 x_1 + n^2 x_2} \\ &=&
%  \frac{|m|^2 + 4 |m| |n| + |n|^2}{m^2 n^2 (|m|+|n|)^4}.
%\end{eqnarray*}
This result is the same as that was found previously for disordered metal rings \cite{danon2010}.

\begin{figure}
\begin{center}
\includegraphics{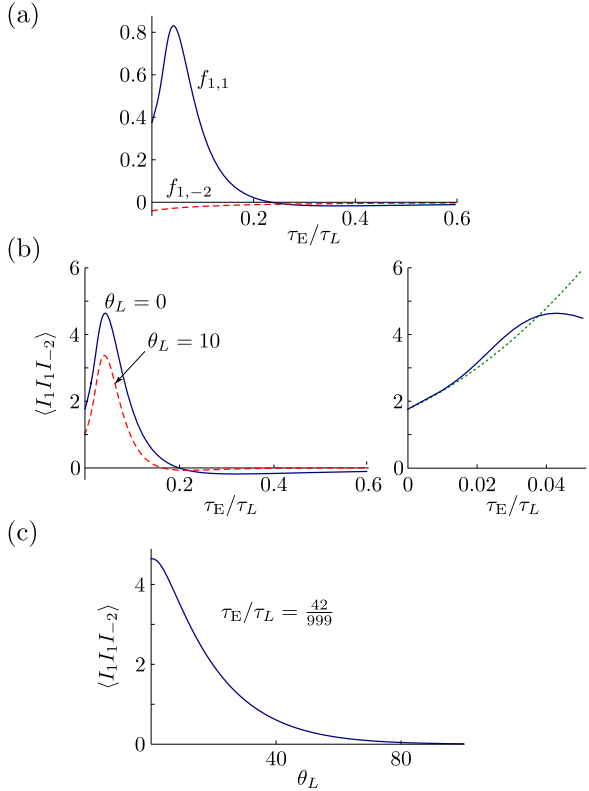}
\caption{\label{fig:f11} (a) Ehrenfest-time dependence of the dimensionless coefficients $f_{1,1}$ (solid) and $f_{1,-2} = f_{-2,1}$ (dashed). (b) Left: The magnitude of the third cumulant $\langle I_1 I_1 I_{-2}\rangle$ in units of $e^3/i\pi^3\tau_L^2\tau_{\rm E}$, shown for zero temperature (solid line) and $\theta_L = 10$ (dashed line). Right: The same zero-temperature result shown for small $\tau_{\rm E}/\tau_L$ (solid line) and the small-$\tau_{\rm E}$ expansion of Eq.\ (\ref{eq:tEexpand}) (dotted line). (c) Temperature dependence of the third cumulant at $\tau_{\rm E}/\tau_L = 42/999 \approx 0.042$, which is close to its maximum at $\theta_L$ = 0.}
\end{center}
\end{figure}

Including a finite Ehrenfest time $\tau_{\rm E}$ into the zero-temperature calculation leads to a modification of the coefficients $f_{n,m}$, which now acquire a dependence on $\tau_{\rm E}/\tau_{L}$. We were not able to perform the integrations over $t_1$ and $t_2$ in closed form at finite Ehrenfest time, but the integrals can be evaluated numerically. The Ehrenfest-time dependence of $f_{1,1}$ and $f_{1,-2}$ is shown in Fig.\ \ref{fig:f11}a and the resulting cumulant $\langle I_1 I_1 I_{-2} \rangle$ is plotted in Fig.~\ref{fig:f11}b in units of $e^3/i\pi^3\tau_L^2\tau_{\rm E}$ (solid line). Two remarkable observations are in place: (i) For moderate but still small values of $\tau_{\rm E}/\tau_{L}$, the inclusion of a finite Ehrenfest time leads to a rather significant enhancement of the non-Gaussian fluctuations. (ii) For larger values of $\tau_{\rm E}/\tau_{L}$ the cumulant changes sign.

In the physically relevant limit of small $\tau_{\rm E}/\tau_{L}$, contributions with $t_1 < \tau_{\rm E}$ or $t_2 < \tau_{\rm E}$ are exponentially small in the large parameter $\tau_L/\tau_{\rm E}$, so that it is sufficient to consider the integral for times $t_{1,2} > \tau_{\rm E}$, for which one can take Eq.\ (\ref{eq:Fresult1})--(\ref{eq:Fresult3}) with $w=0$. The result of a series expansion in the small parameter $\tau_{\rm E}/\tau_{L}$ then yields
\begin{equation}
  f_{n,m} = \sum_{k=0}^{\infty}
  \frac{F_k(n,m) (\tau_{\rm E}/\tau_{L})^k}{(mn)^{k+1}|m n|(|m|+|n|)^{2k + 4}},
  \label{eq:tEexpand}
\end{equation}
where the first three coefficients $F_k(n,m)$ are
\begin{eqnarray*}
  F_0(n,m) &=& m^2 + 4 |mn| + n^2, \\
  F_1(n,m) &=& 16 (m^4 + 6 |m^3 n| 
  \nonumber \\ && \mbox{} + 15 m^2 n^2 + 6 |m n^3| + n^4), \\
  F_2(n,m) &=& 240 (m^6 + 8 |m^5 n|  \nonumber \\ && \mbox{} + 28 m^4 n^2  + 56 |m^3 n^3|  \nonumber \\ && \mbox{}+ 28 m^2 n^4 + 8 |m n^5| + n^6).
\end{eqnarray*}
In Fig.~\ref{fig:f11}b (right plot) we show the cumulant $\langle I_1 I_1 I_{-2}\rangle$ resulting from this second-order expansion (dotted line) together with the full numerical solution (solid line).

For temperature $T \gg \hbar/\tau_L$ we can perform the integrals over $t_1$ and $t_2$ using a saddle-point approximation. In the limit of small Ehrenfest times $\tau_{\rm E} \ll \tau_L$ one finds
\begin{eqnarray}
  \lefteqn{\langle I_n I_m I_{-n-m} \rangle} \nonumber \\ &=&
  \frac{e^3 T^3 \tau_{L}mn(n+m)}{i\tau_{\rm H} \hbar^3}
  \\ \nonumber  && \mbox{} \times
  (g_{m,n} + g_{n,-m-n} + g_{m,-m-n}),
\end{eqnarray}
with $\theta_L \equiv 2 \pi T \tau_L / \hbar \gg 1$ and
\begin{equation}
  g_{m,n} = {\rm sign}(mn) e^{-(|m|+|n|) \sqrt{\theta_L}}.
\end{equation}
We can extend this result to finite Ehrenfest time, yielding
\begin{equation}
  g_{m,n}(\tau_{\rm E}) = {\rm sign}(mn) e^{-(|m|+|n|) \sqrt{\theta_L}} e^{2\,{\rm sign}(mn)\theta_{\rm E}},
  \label{eq:gte}
\end{equation}
with $\theta_{\rm E} \equiv 2\pi T \tau_{\rm E}/\hbar \ll \theta_L$.
% This result is only valid if the temperature is small enough so that the typical times $t_1$ and $t_2$ are not both smaller than $\tau_{\rm E}$. In terms of the parameters in (\ref{eq:gte}) this means that one must have $2\theta_{\rm E} \ll \sqrt \theta_L$, which guarantees that for any $n,m$ the suppression $\sim e^{-\sqrt \theta_L}$ dominates. 
Again, as in the zero-temperature case, at finite temperatures a finite Ehrenfest time can actually lead to an {\em increase} of the third cumulant.

For general temperature, we have to evaluate Eq.~(\ref{eq:Innn}) numerically. In the left panel of Fig.~\ref{fig:f11}b we show the cumulant $\langle I_1 I_1 I_{-2} \rangle$ as a function of $\tau_{\rm E}/\tau_L$ at $\theta_L=10$ (dashed line). The qualitative behavior of the cumulant is the same as at zero temperature: Initially, its magnitude increases, until it reaches a maximum at small but finite $\tau_{\rm E}/\tau_L$. At longer $\tau_{\rm E}$ it decreases again, eventually changing sign at $\tau_{\rm E}/\tau_L \approx 0.2$. In Fig.~\ref{fig:f11}c we show the temperature dependence of $\langle I_1 I_1 I_{-2} \rangle$ at $\tau_{\rm E}/\tau_L = 42/999 \approx 0.042$, which is close to the position of the zero-temperature maximum.

\section{Diagonal contribution to the third cumulant}
\label{sec:4a}

In addition to the off-diagonal contributions to the third cumulant that were discussed in Sec.\ \ref{sec:4}, there are also diagonal contributions that involve orbit repetitions. Such contributions are usually neglected in a semiclassical analysis, because they are suppressed with a factor $e^{-\lambda t}$, where $t$ is the period of the (primitive) orbit and $\lambda$ the Lyapunov exponent. Since the period of typical orbits that encircle the ring $\tau_{L} \gg 1/\lambda$, one argues that such contributions can safely be neglected. However, diagonal contributions do not involve the inverse phase-space volume, so that they lack the factor $\tau_L/\tau_{\rm H}$ that sets the smallness of the off-diagonal contributions such as Eq.\ (\ref{eq:Kresult}).

\begin{figure}
\begin{center}
\includegraphics{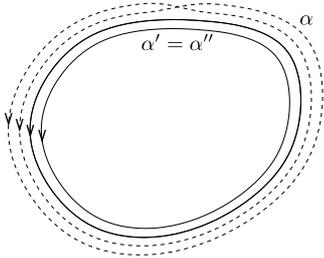}
\caption{\label{fig:loop} Trajectories $\alpha$, $\alpha'$, and $\alpha''$ contributing to the expectation value $\langle I_{1} I_{1} I_{-2} \rangle$. The trajectory $\alpha$ is a non-primitive orbit, consisting of a twofold repetition of $\alpha'=\alpha''$. The trajectories $\alpha'$ and $\alpha''$ each wind once around the ring.}
\end{center}
\end{figure}

The leading such diagonal contribution (for given $m$ and $n$) requires the two short orbits $\alpha'$ and $\alpha''$ to be $n$-fold and $m$-fold repetitions of a primitive orbit $\alpha_0$ with unit winding number, whereas the $\alpha$ is the of the $(m+n)$-fold
repetition of the same orbit, with $m$, $n > 0$. The case $n=m=1$, for which $\alpha'=\alpha''=\alpha_0$, is illustrated in Fig.\ \ref{fig:loop}. For such a diagonal contribution one has a {\em single} sum over orbits,
\begin{eqnarray}
  \lefteqn{ \langle I_n I_m I_{-m-n} \rangle_{\rm c}^{\rm d}} \\ &=& \nonumber
  - \frac{i e^3}{4 \pi^3 \hbar^3}
%  \nonumber \\ && \mbox{} \times
  \sum_{\alpha_0}
  C_{n,m}(t_{\alpha_0})
%  \frac{(\pi T)^3}{\sinh(n \pi t_{\alpha_0} T/\hbar) \sinh(m \pi t_{\alpha_0} T/\hbar)}      }
%  \frac{
  \delta_{n_{\alpha_0},1} |A_{\alpha_0}|^{2n+2m},
%}{\sinh((m+n) \pi t_{\alpha_0} T/\hbar)},
\end{eqnarray}
where we used that we may set $A_{\alpha'} = A_{\alpha_0}^n$, 
$A_{\alpha''}=A_{\alpha_0}^m$, and $A_{\alpha}=A_{\alpha_0}^{m+n}$. We abbreviated
\begin{eqnarray}
  C_{n,m}(t) &=& \frac{(\pi T)^2}{\sinh(n \pi t T/\hbar) \sinh(m \pi t T/\hbar)}
  \nonumber \\ && \mbox{} \times
  \frac{\pi T}{\sinh[(m+n) \pi t T/\hbar]}.
\end{eqnarray}
For uniformly hyperbolic dynamics one has $|A_{\alpha_0}|^2 = 1/2 \sinh(\lambda t_{\alpha_0}) \approx e^{- \lambda t_{\alpha_0}}$ for $\lambda t_{\alpha_0} \gg 1$ \cite{haake1991}. This then gives 
\begin{eqnarray}
  \lefteqn{\langle I_n I_m I_{-m-n} \rangle_{\rm c}^{\rm d}} \nonumber \\ &=&
  - \frac{i e^3}{4 \pi^3 \hbar^3} \sum_{\alpha_0}  C_{n,m}(t_{\alpha_0})
  \nonumber \\ && \mbox{} \times
%  \frac{(\pi T)^3}{\sinh(n\pi t_{\alpha_0} T/\hbar) \sinh(m\pi t_{\alpha_0} T/\hbar)     }
%  \frac{
  \delta_{n_{\alpha_0},1} |A_{\alpha_0}|^2 e^{-(m+n-1)\lambda t_{\alpha_0}}.
%}{\sinh((m+n) \pi t_{\alpha_0} T/\hbar)}.
\end{eqnarray}
Performing the remainder of the calculation as in Sec.\ \ref{sec:3}, one finds that the diagonal contribution to the third cumulant of the persistent current reads
\begin{eqnarray}
  \lefteqn{\langle I_n I_m I_{-n-m} \rangle_{\rm c}^{\rm d}}
  \nonumber \\  &=&
    - \frac{i e^3}{4 \pi^3 \hbar^3}
  \int \frac{dt}{t}\, C_{n,m}(t)  \sqrt{\frac{\tau_{L}}{4 \pi t}} 
  e^{-\tau_{L}/4 t}
  \nonumber \\ && \mbox{} \times
%  \frac{(\pi T)^3 e^{-(m+n-1)\lambda t}}{t \sinh (n \pi t T/
%\hbar)\sinh(m\pi t T/\hbar) \sinh((m+n) \pi t T/\hbar) }.
  e^{-(m+n-1)\lambda t}.
% \nonumber \\
\end{eqnarray}
For temperatures $T \ll \hbar \sqrt{\lambda/\tau_{L}}$ one then finds an essentially temperature-independent diagonal contribution to the third cumulant of the persistent-current fluctuations, 
\begin{eqnarray}
  \lefteqn{\langle I_n I_m I_{-n-m} \rangle_{\rm c}^{\rm d}} \\ &=& \nonumber
  -2i\frac{e^3 \lambda^{3/2}}{\pi^3 \tau_{L}^{3/2}}\frac{(m+n-1)^{3/2}}{mn(m+n)} 
%  \nonumber \\ && \mbox{} \times
e^{-\sqrt{(m+n-1)\lambda \tau_{L}}}.
\end{eqnarray}

At zero temperature all diagonal contributions from orbit repetitions are smaller than the off-diagonal contributions if the condition 
\be
  \frac{\tau_{\rm H}}{\tau_L} \ll \frac{e^{\sqrt{(m+n-1)
  \lambda \tau_L}}}{(\lambda \tau_L)^{3/2}}
\ee
is met. This condition can also be rephrased in terms of the Ehrenfest time $\tau_{\rm E}$, using $\tau_{\rm E} \sim \lambda^{-1} \ln(\tau_{\rm H}/\tau_L)$, as
\begin{equation}
  \tau_{\rm E} \lesssim \frac{(n+m-1)\tau_{L}}{\ln(\tau_{\rm H}/\tau_L)}.
  \label{eq:tauEbound}
\end{equation}
Given the intrinsic smallness of the Ehrenfest time, this condition is easily met. However, to see a nontrivial Ehrenfest-time dependence of the persistent current fluctuations, the Ehrenfest time needs to be a finite fraction of $\tau_L$, see Sec.\ \ref{sec:4}, and the condition (\ref{eq:tauEbound}) effectively limits the applicability of the results as shown in Fig.\ \ref{fig:f11} to the range $\tau_{\rm E}/\tau_{L} \ll 1$, where the expansion (\ref{eq:tEexpand}) is valid.
%This condition is always met in the limit $\tau_{\rm E}/\tau_{L} \to 0$ at fixed ratio $\tau_{\rm H}/\tau_L$; however, it may still be met for finite but small Ehrenfest times, provided $\tau_{\rm H}/\tau_L$ is not too large. 

At finite temperatures $T \gtrsim \hbar/\tau_L$ the off-diagonal contribution is strongly suppressed and the diagonal contribution quickly takes over. This reflects the large difference in typical orbit durations for the
off-diagonal and diagonal contributions: For the off-diagonal contribution, the typical orbit duration $\sim \tau_L$ at zero temperature, so that temperature starts to suppress this contribution for $T \gtrsim \hbar/\tau_L$. The typical duration of orbits contribution to the diagonal contribution is $\sim \sqrt{\tau_L/\lambda}$, which explains the relative insensitivity of this contribution to temperature.

\section{Discussion and conclusion}
\label{sec:5}

The distribution of the persistent current in a mesoscopic ring is Gaussian, with small non-Gaussian corrections. Here we have
presented a semiclassical calculation of the leading non-Gaussian correction, described by the three-point correlation function $K = \langle I(\phi_1)I(\phi_2)I(\phi_3) \rangle_{\rm c}$. 
In agreement with previous work for disordered metal rings \cite{danon2010,houzet2010}, we found here that at small temperatures  $K \sim e^3/\tau_L^2\tau_{\rm H} = e^3/g\tau_L^3$, where $g = \tau_{\rm H}/\tau_{L}$ is the dimensionless conductance of the ring, $\tau_L$ the diffusion time, and $\tau_{\rm H}$ the Heisenberg time. The semiclassical approach also contains information on the role of the Ehrenfest time in such a ring, and we showed that for small but finite $\tau_{\rm E}/\tau_L$ the magnitude of the non-Gaussian corrections is enhanced by a numerical factor, before it is suppressed in the limit of large $\tau_{\rm E}/\tau_{L}$. 

The fact that the three-point correlation function initially increases with increasing Ehrenfest time is remarkable, since a finite Ehrenfest time usually suppresses quantum interference effects. However, it is not without precedent: The conductance fluctuations in a chaotic cavity are Ehrenfest-time independent \cite{tworzydlo2004,jacquod2004}, whereas the conductance fluctuations in a quasi-one dimensional Lorentz gas are larger in the limit of large Ehrenfest time than in the limit of zero Ehrenfest time \cite{brouwer2007b}. The same applies to the variance of the current pumped through a chaotic cavity with a periodic modulation of its shape \cite{rahav2006c}. Just as in the present case, the conductance fluctuations or the variance of the pumped current contain contributions from closed loops, and it is this type of correction that can in principle be enhanced by Ehrenfest-time corrections.

We have also identified a second semiclassical contribution to $K$, which involves a diagonal orbit sum with non-primitive periodic orbits. Non-primitive orbits are usually neglected in semiclassical approaches, because their contribution is exponentially suppressed in comparison to contributions from
primitive orbits. Nevertheless, for the three-point correlation of the persistent current, such diagonal contributions become dominant in the limit of large Ehrenfest times and/or temperatures. A related (but not identical) effect appears for conductance fluctuations in a chaotic quantum dot, where ``classical fluctuations'' become important for large Ehrenfest times \cite{tworzydlo2004}.

The small magnitude of the non-Gaussian fluctuations turns its
measurement into a considerable challenge, even with state-of-the-art techniques \cite{bleszynski2009,castellanos-beltran2013}. For disordered metal rings, the conditions for measuring non-Gaussian corrections to the distribution are most favorable if its dimensionless conductance $g$ is not too large,
since one needs to average over at least $\sim g^2$ statistically independent samples to be able to distinguish the three-point function from the Gaussian (second-order) fluctuations \cite{castellanos-beltran2013}. For Ehrenfest-time-related corrections to become relevant merely 
making $g$ small is not the solution, since a small dimensionless conductance $g$ also implies a small $\tau_{\rm E}$.

We gratefully acknowledge discussions with Alexander Altland, Fritz Haake, Sebastian M\"uller, and Felix von Oppen.
This work is supported by the Alexander von Humboldt Foundation in the framework of the Alexander von Humboldt Professorship, endowed by the Federal Ministry of Education and Research (PWB).
\bigskip

\bibliographystyle{elsarticle-num}

%\bibliography{/home/brouwer/grant/2012/refs}

\end{document}